\documentclass[aps,prb,twocolumn,groupedaddress]{revtex4}
\usepackage{txfonts}
\usepackage{amssymb}
\usepackage{graphicx}
\usepackage{color}
\usepackage[colorlinks,citecolor=blue, linkcolor=blue, urlcolor=black, dvipdfm]{hyperref}

\begin{document}

\title{Surface and bulk electronic structures of LaOFeAs studied by angle resolved photoemission spectroscopy}

\author{L. X. Yang$^1$, B. P. Xie$^1$ }\email[]{bpxie@fudan.edu.cn}
\author{Y. Zhang$^1$, C. He$^1$, Q. Q. Ge$^1$, X. F. Wang$^2$, X. H. Chen$^2$}
\author{M. Arita$^3$, J. Jiang$^3$, K. Shimada$^3$, M. Taniguchi$^3$, I. Vobornik$^4$, G. Rossi$^{4,5}$, J. P. Hu$^6$, D. H. Lu$^7$, Z. X. Shen$^7$, Z. Y. Lu$^8$}
\author{D. L. Feng$^1$}\email[]{dlfeng@fudan.edu.cn}

\affiliation{$^1$ State Key Laboratory of Surface Physics, Department of Physics, and Advanced Materials Laboratory, Fudan University, Shanghai 200433, P. R. China}

\affiliation{$^2$Hefei National Laboratory for Physical Sciences at
Microscale and Department of Physics, University of Science and Technology of China, Hefei, Anhui 230026, P. R. China}

\affiliation{$^3$Hiroshima Synchrotron Radiation Center and Graduate School of Science, Hiroshima University, Hiroshima 739-8526, Japan}

\affiliation{$^4$CNR-INFM, TASC Laboratory AREA Science
$Park-Basovizza$, 34012, Trieste, Italy}

\affiliation{$^5$Dipartimento di Fisica, Universita¡ä di Modena e
Reggio Emilia, Via Campi 213/A, I-41100 Modena, Italy}

\affiliation{$^6$Department of physics, Purdue University, West
Lafayette, Indiana 47907, USA}

\affiliation{$^7$Stanford Synchrotron Radiation Lightsource, SLAC
National Accelerator Laboratory, 2575 Sand Hill Road, Menlo Park, CA
94025, USA}

\affiliation{$^8$Department of Physics, Renmin University of China,
Beijing 100872, P. R.  China. }

\date{\today}

\begin{abstract}

The electronic structure of LaOFeAs, a parent compound of
iron-arsenic superconductors, is studied by angle-resolved
photoemission spectroscopy. By examining its dependence on photon
energy, polarization, sodium dosing and the counting of Fermi
surface volume, both the bulk and the surface contributions are
identified. We find that a bulk band moves toward high binding
energies below structural transition, and shifts smoothly across the
spin density wave transition by about 25~meV. Our data suggest the
band reconstruction may play a crucial role in the spin density wave
transition, and the structural transition is driven by the short
range magnetic order. For the surface states, both the
LaO-terminated and FeAs-terminated components are revealed. Certain
small band shifts are verified for the FeAs-terminated surface
states in the spin density wave state, which is a reflection of the
bulk electronic structure reconstruction. Moreover, sharp
quasiparticle peaks quickly rise at low temperatures, indicating of
drastic reduction of the scattering rate. A kink structure in one of
the surface band is shown to be possibly related to the
electron-phonon interactions.

\end{abstract}


\maketitle


\section{Introduction}

The discovery of superconductivity in LaO$_{1-x}$F$_x$FeAs has
declared the advent of iron-based high temperature superconductors
(Fe-HTSC's). \cite{JACS, XHChenNature, GFChenCe} Resembling the
cuprates, superconductivity emerges from the anti-ferromagnetic
(AFM) ordered ground state upon proper doping.\cite{XHChenNature,
HosonoNature, GFChenCe, MXuTh} The intimate relationship between
superconductivity and magnetism makes it critical to study the
magnetic fluctuations in the parent compounds. On the other hand,
the isotope effects, \cite{ChenISO} along with the ubiquitous
co-occurrence of the structural and magnetic transitions, allude to
the relevance of the lattice degree of freedom for the Fe-HTSC
physics. Therefore, it is indispensable to reveal the origin of the
structural transition and its effects on the electronic structure as
well as the role of phonons for a comprehensive understanding of the
Fe-HTSC's. After intensive research, the record superconductivity
transition temperature ($T_c$) of Fe-HTSC's is still held by the
$Ln$O$_1-x$F$_x$FeAs ($Ln$=La, Sm, and Ce etc, the so called ``1111"
series.) systems.\cite{RecordTc, RecordTc2} Moreover, the divergence
of the structural and spin density wave transitions in parent
compounds of the ``1111" series provides a rare opportunity to
unveil these crucial problems.\cite{PCDaiNature}

Angle-resolved photoemission spectroscopy (ARPES) has been employed
to study the electronic structure of various Fe-HTSC's, revealing
the electronic structure, superconducting gap and the electron-boson
coupling.\cite{DHLuNature, DHLuPhC, HDing, HDing2, LXYang, YZhang,
MYi, ZX_PM, BZhou} However, due to the covalent La-O bonding,
$Ln$OFeAs exposes a polar surface with charge redistribution after
cleavage.\cite {DHLuNature, Kaminski, XJZhouCe} Indeed, a recent
detailed band calculation of LaOFeAs shows that the electronic
structure of the surface deviates strongly from that of the bulk,
and there are two types of surfaces: the LaO-terminated surface and
the FeAs-terminated one.\cite{surface} ARPES is essentially a
surface probe, therefore, the bulk electronic structure of $Ln$OFeAs
is still not definite. This prevents the understanding of this very
first and probably the highest-$T_c$ series of iron pnictides, which
in turn hampers the construction of a global picture of electronic
structure in iron pnictides.

In this Article, we report the ARPES measurements of the electronic
structure of LaOFeAs, a parent compound of the ``1111" series of
iron pnictides. By carefully conducting photon energy dependence and
Na-dosing dependence studies on the complicated electronic
structure, we could distinguish states from the bulk and the
surface. As a result, the nature of  the  spin density wave (SDW)
state is exposed by the bulk bands. Similar to BaFe$_2$As$_2$ and
other members in the ``122" series, no energy gap related to nesting
is observed at the Fermi surfaces \cite{LXYang, YZhang, MYi, BZhou}.
Instead, a band at high binding energies shifts down by about
25~meV, starting from the structural transition temperature. This
has been observed recently in NaFeAs by the authors,\cite{CHe} where
the fluctuating magnetic order was shown to appear at the structural
transition and drive the structural order. Our results on LaOFeAs
further suggest that the SDW picture established in the ``122" and
``111" series applies in the ``1111" series as well. The
reconstruction (or shift) of the band structure effectively saves
the energy, and thus plays a crucial role in driving the SDW
transition. Furthermore,  we observed drastically rising coherent
quasiparticles, which indicates that scattering is strongly
suppressed at low temperatures. A kink in dispersion related to
possible surface electron-phonon interactions is observed as well.

\begin{figure}[t]
\includegraphics[width=8.5cm]{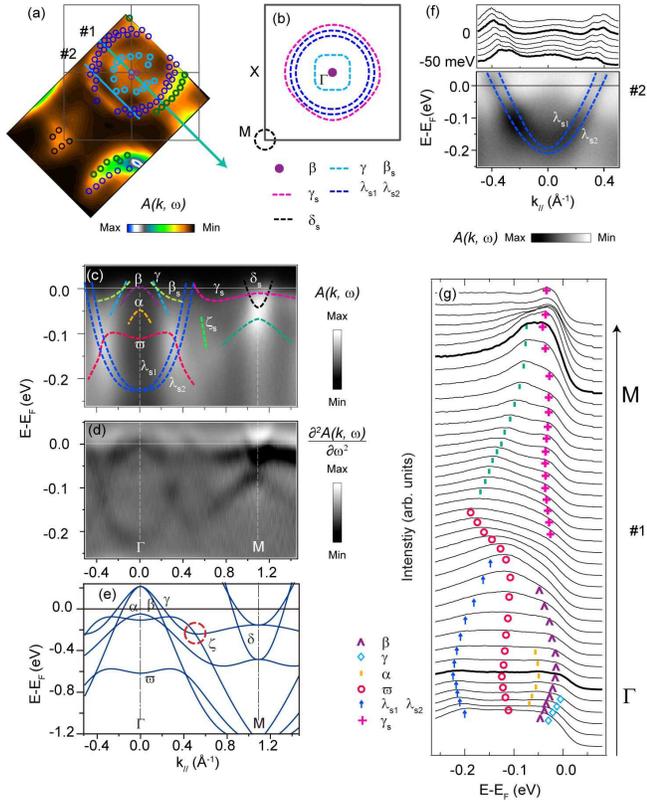}
\caption{(color online) Fermi surface mapping and band structure of
LaOFeAs along $\Gamma$-$M$ in normal state (170~K). (a) ARPES
intensity map in the Brillouin zone (BZ) integrated over $\pm$ 5~meV
around the Fermi energy ($E_F$), overlaid by the Fermi crossings.
(b) The Fermi surfaces are constructed by tracking the Fermi
crossings. (c) Photoemission intensity plot along $\Gamma$-$M$ (cut
1 in panel (a)). The dashed curves are the guides to eyes obtained
by tracking the local minimum locus of second derivative of the raw
data with respect to energy as shown in panel (d). (e) The
calculated bulk electronic structure of LaOFeAs along $\Gamma$-$M$
based on density functional theory. (f) The momentum distribution
curves (MDC's) (upper panel) and photoemission intensity plot
divided by the resolution convoluted Fermi-Dirac function (lower
panel) along cut 2 as marked in panel (a). (g) The corresponding
energy distribution curves (EDC's) along $\Gamma$-$M$. Data were
taken using 24~eV circularly polarized photons.}\label{normal}
\end{figure}

\begin{figure}[t]
\includegraphics[width=8.5cm]{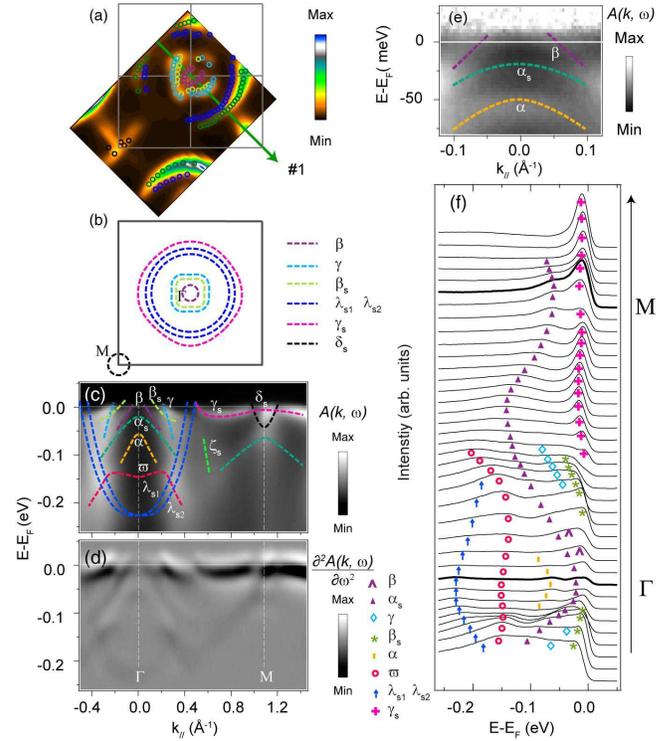}
\caption{(color online) Fermi surface mapping and band structure
along $\Gamma$-$M$ of LaOFeAs in the SDW state (10~K). (a) ARPES
intensity map in the BZ integrated over $\pm$ 5~meV around $E_F$,
overlaid by the Fermi crossings. (b) The Fermi surfaces are
constructed by tracking the Fermi crossings. (c) Photoemission
intensity plot along $\Gamma$-$M$ (cut 1 in panel (a)). The dashed
curves are the guides to eyes obtained by tracking the local minimum
locus of second derivative of the raw data with respect to energy as
shown in panel (d). (e) The photoemission intensity plot after each
MDC is normalized by its integrated weight. (f) The corresponding
EDC's along $\Gamma$-$M$. Data were taken using 24~eV circularly
polarized photons.}\label{sdw}
\end{figure}

\section{Experimental}

High quality LaOFeAs single crystals were synthesized by NaAs-flux
method as described elsewhere. \cite{Growth} Resistivity data
confirmed the SDW transition at $T_N$=138~K.\cite {PCDaiNature}
ARPES measurements were conducted at Beamline 9 of Hiroshima
synchrotron radiation center (HSRC) with circularly polarized
photons and a Scienta R4000 electron analyzer. The Polarization and
photon-energy dependence measurements were performed at Beamline 1
of HSRC and the APE Beamline in Elettra synchrotron light source.
The energy resolution is 9~meV at Beamline 9, and 20~meV at Beamline
1 and APE, respectively. The overall angular resolution is about
0.3$\textordmasculine$. All samples were cleaved \textit{in situ}
and the ARPES measurements were carried out under ultra-high-vacuum
better than 3.0$\times$10$^{-11}$~mbar at Beamline 9 and
5$\times$10$^{-11}$~mbar at Beamline 1 of HSRC. Data were taken
within 6~hours after cleavage to minimize the aging effect.

The energy of photons in the experiments ranges from 19 to 64~eV,
which gives an electron escape  depth of about 5-10 $\AA$  according
to the universal curve of the electron mean free path in solid. This
corresponds to the first two or three layers of LaO or FeAs plane in
LaOFeAs.  An inner potential of 16~eV is chosen to determine the
$k_z$ of the high symmetry points of the Brillouin zone in the
photon energy dependence studies.

\section{Data Analysis and Discussion}

\subsection{Normal state electronic structure}

The electronic structure in normal state (170~K) are presented in
Fig.~\ref{normal}. Photoemission intensities are integrated over a
[-5~meV, +5~meV] window at $E_F$ to acquire Fermi surface as shown
in Figs.~\ref{normal}(a) and \ref{normal}(b). The observed Fermi
surface consists of three hole pockets, two electron pockets and a
tiny patch like feature around the $\Gamma$ point as well as one
electron pocket around $M$. This is very different from the
calculated bulk electronic structure and what is observed in other
iron pnictides.\cite{IIMazin, ZYLu} The band structure as indicated
by the dashed curves in Fig.~\ref{normal}(c) is resolved by tracking
the local minimum locus in the second derivative of the ARPES
intensity plot with respect to energy (Fig.~\ref{normal}(d)) and
confirmed by the peaks in the corresponding energy distribution
curves (EDC's) as shown in Fig.~\ref{normal}(g). The top of the
$\beta$ band touches the $E_F$ and forms a Fermi patch near
$\Gamma$. The Fermi crossings of $\gamma$ and $\beta_s$  nearly
coincide, forming two hole pockets of the Fermi surface. Two
parabolic electron bands $\lambda_{s1}$ and $\lambda_{s2}$
contribute two electron Fermi pockets around $\Gamma$. A weak but
resolvable feature $\alpha$ shows up with its top at about 50~meV
below $E_F$. The flat $\gamma_s$ band contributes one large hole
pocket around $\Gamma$ which is much larger than the corresponding
bulk Fermi surface in the calculation. Around the $M$ point, only
one electron band could be clearly resolved. Note that the petal
like feature around $M$ in Fig.~\ref{normal}(a) stems from the
remnant spectral weight of the flat $\gamma_s$ band, instead of
Fermi crossings.

The overall measured electronic structure of LaOFeAs deviates
strongly from that of other pnictides and the calculations as shown
in Fig.~\ref{normal}(e).\cite{LXYang, IIMazin, ZYLu, SinghCalc} In
total, we observed 9 bands and 5 Fermi surface sheets around
$\Gamma$ (actually, another hole like band $\alpha_s$ could be
resolved around $\Gamma$ at low temperature as shown below.) instead
of 6 bands and 2 Fermi surface sheets in the bulk band calculations.
This complication can be qualitatively explained by the recent
calculations of the surface and bulk band structures of
LaOFeAs.\cite{surface}  Because of the strong La-O or Fe-As covalent
bonding, the  surface is either an FeAs or a LaO plane. The polarity
of the surface causes  significant lattice relaxation and charge
redistribution to minimize its static electric energy and thus
changes the electronic structure dramatically. \cite{DHLuNature,
DHLuPhC}  As a result, the FeAs-terminated surface is
electron-deficient, while the LaO-terminated surface is
electron-excess. As ARPES probes only 5$\sim$10~\AA with the photon
energies exploited here, the measured electronic structure are
contributed from the first 2 or 3 LaO or FeAs planes. Because the
two types of surfaces could coexist on a single cleaved sample, and
the FeAs plane underneath the LaO surface is different from the FeAs
surface plane, they conspire to construct the complicated band
structure in the experimental data.

The surface band calculations (Ref.\onlinecite{surface}) clearly
predicted that the excess electrons in the surface LaO plane reside
in the La $5d+6s$ states, and give a large electron pocket around
$\Gamma$. However, instead of one pocket, we have observed two large
electron pockets ( $\lambda_{s1}$ and $\lambda_{s2}$), which could
be resolved more clearly in Fig.~\ref{normal}(f). Since the observed
two bands mimic the spin-orbital splitting of the Au(111) surface
states, the difference between our data and the predicted ones could
be caused by the fact that spin-orbital coupling was not included in
the calculation, while the spin-orbital coupling of $5d$ electrons
is not negligible.\cite{CKim} For comparison, the momentum splitting
of $\lambda_{s1}$ and $\lambda_{s2}$ is about 0.03$\AA^{-1}$ and the
Rashba energy is merely $\sim$1~meV. Correspondingly,  the Rashba
parameter $\alpha_R$, \textit{i.e.}, the ratio of the Rashba energy
to the momentum splitting, is about 0.023~eV$\AA^{-1}$. This value
is much smaller than that of Au: $\alpha_R$=0.33~eV$\AA^{-1}$,
\cite{AuSO} but comparable with that of the two dimensional
electronic system in the InGaAs/InAlAs heterostructure. \cite{HSSO}
Furthermore, we note that the occupied bandwidth of $\lambda_{s1}$
and $\lambda_{s2}$ is about 0.22~eV, which is similar to the
calculation (0.2~eV), indicating weak correlations in the LaO layer,
in contrast to the FeAs bands.\cite{FChen}

\begin{figure*}
\includegraphics[width=16cm]{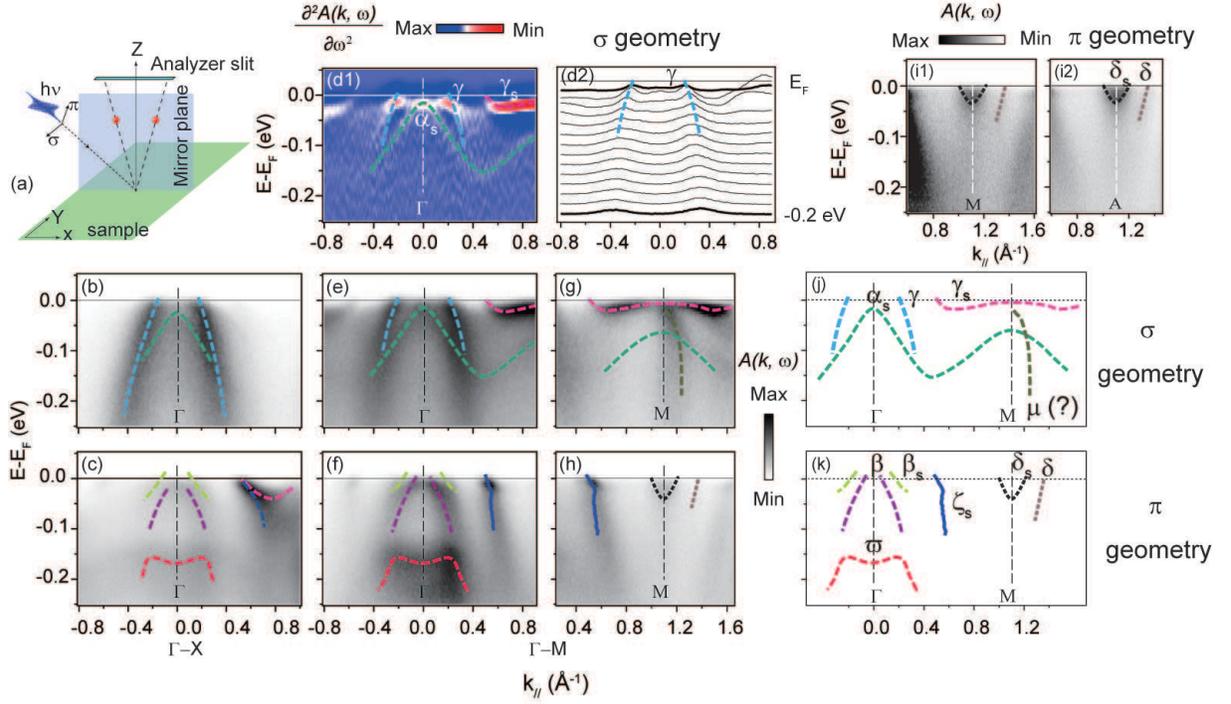}
\caption{(color online) Polarization dependent measurements along
high symmetry directions. (a) Schematic diagram of two types of
experimental setups. (b)-(c) ARPES intensity plot along $\Gamma$-$X$
(d1) The second derivative of ARPES intensity plot with respect to
energy and (d2) the corresponding MDC's along $\Gamma$-$M$. (e)-(h)
ARPES intensity plot along $\Gamma$-$M$. (i1)-(i2) ARPES intensity
plot around the $M$ and $A$ points. The contrast of the images is
adjusted to show the $\delta$ band more clearly. (j) and (k)
summarize the band structure extracted for the corresponding
polarizations. The dashed curves in the photoemission intensity
plots are the guides to eyes. The geometry of experimental setups
are marked on the top or right of the panels. The corresponding
photon energies for the $\Gamma$, $M$ and $A$ points are 59~eV,
43~eV and 33~eV respectively to reach the right $k_z$. Data were
taken at 10~K.}\label{p-dep}
\end{figure*}

\subsection{SDW state electronic structure}

Figure~\ref{sdw} displays the corresponding electronic structure in
the SDW state. For the Fermi surface as shown in Figs.~\ref{sdw}(a)
and \ref{sdw}(b), the tiny patch like feature at $\Gamma$ evolves
into a small hole pocket. The Fermi crossings of the $\gamma$ and
$\beta_s$ bands does not coincide any more, forming two hole pockets
around the $\Gamma$ point. The size of $\gamma$ pocket expands about
70\% at low temperature. There is little change of the other Fermi
surface sheets. The band structure along $\Gamma$-$M$ is shown in
the ARPES intensity plot (dashed curves in Fig.~\ref{sdw}(c)) and
its second derivative with respect to energy (Fig.~\ref{sdw}(d)). An
additional hole like band ($\alpha_s$) with its band top at about
-20~meV could be resolved in the Fig.~\ref{sdw}(e). The reason that
$\alpha_s$ was not resolved at high temperatures might be due to
thermal broadening. The splitting of the $\lambda_{s1}$ and
$\lambda_{s2}$ bands is independent of temperature, which is
expected for spin orbital splitting. The $\alpha$ and $\varpiup$
bands move about 5~meV and 30~meV downward respectively, saving the
total energy of the system. The $\gamma_s$ band moves about
$\sim$5~meV towards $E_F$, sharpening the petal like feature around
the $M$ point without additional Fermi crossing of the $\gamma_s$
band. No clear variation of the electron band around $M$ is found.
Moreover, no energy gap related to Fermi surface nesting was
observed, similar to the case of the ``122", ``111" and "11" series.
\cite{BZhou,LXYang, YZhang, MYi, CHe, FChen}.

\subsection{Polarization dependence}

The multiband nature of the iron pnictides leads to complex band
structure, and the additional surface states make it even more
complicated in LaOFeAs. Since the band structure shows strong
polarization dependence, we conducted polarization dependent ARPES
measurements to further resolve the band structure.
Fig.~\ref{p-dep}(a) displays the definition of two different
experimental geometries according to the linear polarization of the
incoming photons. The incident beam and the sample surface normal
define a mirror plane. For the $\sigma$ (or $\pi$) experimental
geometry, the electric field of the incident photons is out of (or
in) the mirror plane. The matrix element for the photoemission
process could be described as

$${M_{f,i}^{\bf{k}}\propto{\rm{|}}\langle \Psi _{f}^{\bf{k}}{\rm{|}}\bf{\hat{\varepsilon}}\cdot\bf{r}\rm{|}\Psi _{\it{i}}^{\bf{k}}\rangle
{\rm{|}}^2}$$

Since the final state $\Psi_f$ of photoelectrons could be
approximated by a plane wave with its wave vector in the mirror
plane, $\Psi_f$ is always even in our experimental geometry. In the
$\sigma$ (or $\pi$) geometry, $\ifmmode\hat{\varepsilon}\fi{} \cdot
\textbf{r}$ is odd (or even) with respect to the mirror plane. Thus,
only the odd (or even) component of the initial state $\Psi_i$ could
show up in the photoemission data. \cite{FChen}

The photoemission intensity plot along high symmetry directions are
shown in Figs.~\ref{p-dep}(b)-(c) and (e)-(h). The dashed lines are
the guides to eyes while the solid curves are the fitting of MDC's
peaks. Around the $\Gamma$ point, the $\beta$ and $\beta_s$
($\alpha_s$ and $\gamma$) bands only shows up in $\pi$ ($\sigma$)
geometry of experimental setup, exhibiting their even (odd) nature
with respect to the mirror plane. The flat $\gamma_s$ band is absent
in $\sigma$ ($\pi$) geometry along $\Gamma$-X ($\Gamma$-$M$). It
exhibits different parity in two high symmetry directions. The
$\varpi$ band could be observed only in $\pi$ geometry along both
directions. Its parity is analogous to that of the $d_{z^2}$ band as
predicted in density functional theory.\cite{FChen}
Fig.~\ref{p-dep}(d1) shows that the $\alpha_s$ band disperses from
$\Gamma$ to $M$ and connects smoothly with the hole like band there.
Around $M$ point, the $\delta$ and $\delta_s$ bands only shows up in
the $\pi$ experimental geometry, indicating their even parities. The
contrast of the ARPES intensity plots around $M$ and $A$ are
adjusted to show the $\delta$ band more clearly in
Figs.~\ref{p-dep}(i1) and (i2). Note that we observed a fast
dispersed hole like band $\mu$ around the $M$ point, whose origin is
still not clear yet (Fig.~\ref{p-dep}(g)). We have reproduced the
obtained band structure under different experimental geometries in
Figs.~\ref{p-dep}(j) and (k) to summarize the band structure along
$\Gamma$-$M$.

An interesting result in Figs.~\ref{p-dep}(f) and 3(h) is the kink
like dispersion of the $\zetaup_s$ band, which has been reported
earlier in CeOFeAs as well.\cite {XJZhouCe} The dispersion of the
$\zetaup_s$ band is tracked by fitting its MDC's (the blue solid
curve). A kink does appear at about -30~meV, which might indicate
strong electron-phonon interactions for this particular band.
However, band crossings could cause similar structure. A kink could
be possibly introduced by the hybridization of the $\gamma_s$ and
$\zetaup_s$ bands as highlighted by the red dashed circle in
Fig.~\ref{normal}(e), there might be a band crossing in this
momentum region. Correspondingly, the $\gamma_s$ band should connect
either to the $\gamma$ or the $\beta_s$ band to hybridize with the
$\zetaup_s$ band. However, the $\gamma$ band could be clearly
resolved to disperse quickly downward from the MDC's in
Fig.~\ref{p-dep}(d2), it could not connected to the $\gamma_s$ band.
On the other hand, $\beta_s$ and $\gamma_s$ have the opposite
parities as shown in Figs.~\ref{p-dep}(j) and (k). Based on LDA
calculations and our previous studies of the orbital nature of bands
in BaFe$_{2-x}$Co$_x$As$_2$, \cite{YZhangOrbital} a band should not
change its parity in such a small momentum range. Therefore,
$\beta_s$ and $\gamma_s$ bands could not be from the same band. In
conclusion, the polarization data further confirms that the kink is
not caused by a band crossing, but most likely intrinsic
electron-phonon interactions.

\begin{figure*}
\includegraphics[width=15cm]{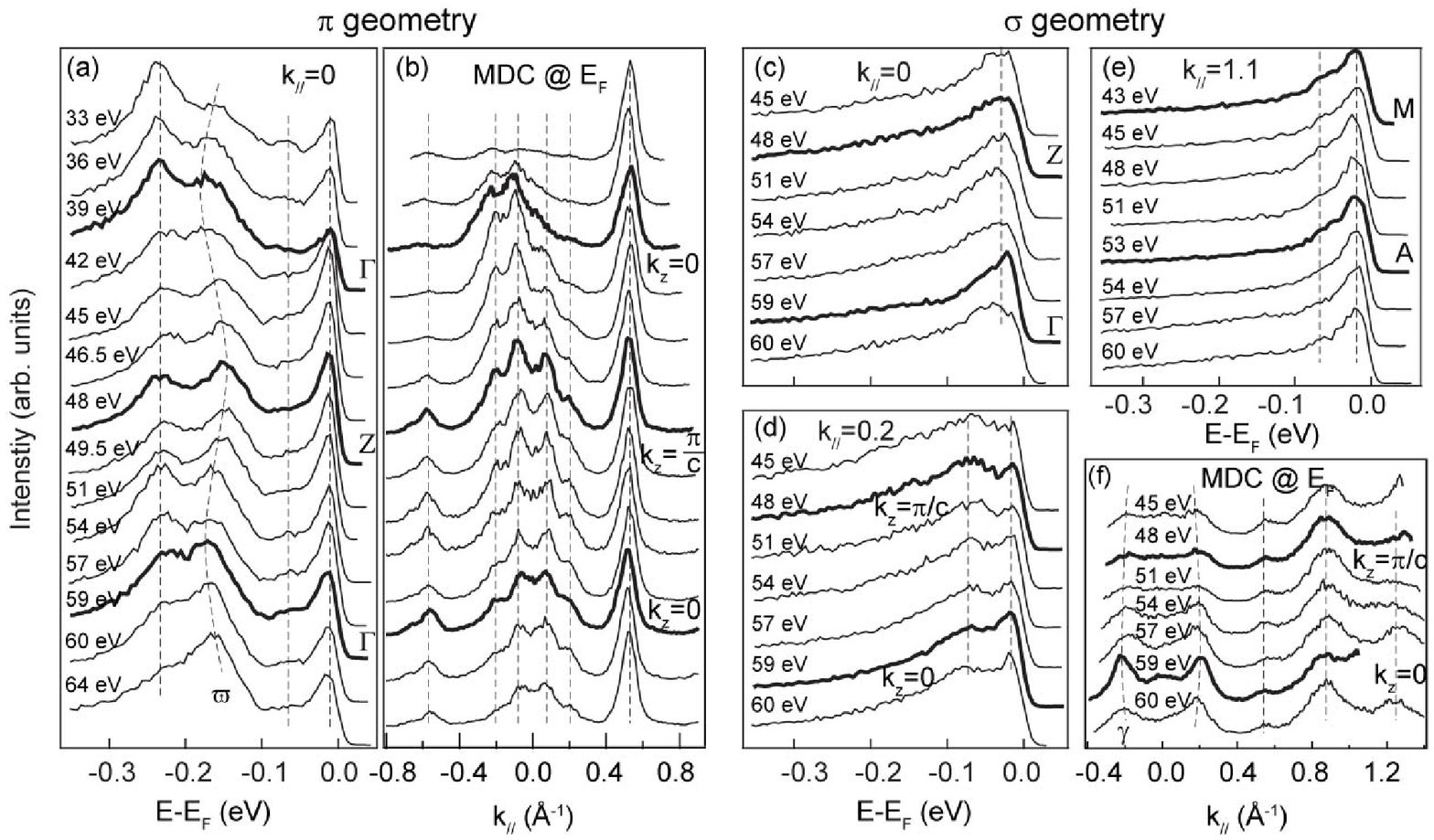}
\caption{(color online) Photon energy dependence of (a) the EDC's at
in-plane momentum $k_{\varparallel}=0$, and (b) the MDC's at $E_F$.
The EDC's at in-plane momentum (c) $k_{\varparallel}=0$, (d)
$k_{\varparallel}=0.2$ $\AA^{-1}$ and (e) $k_{\varparallel}=1.1$
$\AA^{-1}$. (f) the MDC's at $E_F$. Data in panel (a) and (b) were
taken under $\pi$ geometry, while data in panel (c)-(e) were taken
under $\sigma$ geometry of experimental setup. The in-plane
projection are along the $\Gamma$-$M$ direction. Data were taken at
10K.}\label{phE}
\end{figure*}

\begin{figure}[t!]
\includegraphics[width=8.5cm]{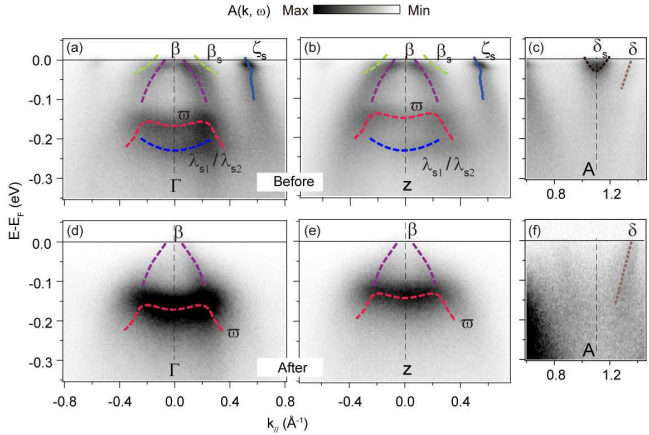}
\caption{(color online) Photoemission intensity plots along (a) the
$\Gamma$-$M$ direction, (b) the Z-A direction and (c) the $Z$-$A$
direction under $\pi$ geometry of experimental setup as in
Fig.~\ref{p-dep}(a). (d)-(f) After the sample was exposed to Na
dosing, data were taken at the same condition as panels a and b
respectively. The dashed curves are the guides to eyes. Data were
taken at 10~K under $\pi$ experimental geometry. The corresponding
photon energies for the $\Gamma$, $Z$ and $A$ points are 59~eV,
48~eV, and 33~eV, respectively.}\label{Na-dose}
\end{figure}

\subsection{Identification of surface and bulk electronic structure: $k_z$ dependence and Na dosing effects}

It is  crucial to identify the bulk and surface bands to reveal the
intrinsic response of the electronic structure to the phase
transitions. Photon-energy dependence measurements were carried out
for this purpose,  because different photon energies correspond to
different  $k_z$, and  only bulk bands could exhibit
$k_z$-dispersion.  Under the $\pi$ geometry of experimental setup,
as shown in Figs.~\ref{phE}(a) and \ref{phE}(b), only the $\varpiup$
band has noticeable $k_z$ dependence as its energy position at
$k_{\varparallel}=0$ changes with the variation of photon energy.
Thus, it should be one of the bulk bands. No clear $k_z$ dependence
is observed for the bands crossing $E_F$ according to the photon
energy dependence of the MDC's at $E_F$ as shown in
Fig.~\ref{phE}(b). For the bands observed under $\sigma$
experimental geometry, only weak $k_z$ dependence could be verified
for the $\gamma$ band from the MDC's at $E_F$ as shown in the
Fig.~\ref{phE}(f). No other bands show noticeable $k_z$ dependence
as shown by the EDC's at $k_{\varparallel}=0$,
$k_{\varparallel}=1.1$\AA and $k_{\varparallel}=0.2$\AA
(Figs.~\ref{phE}(c)-\ref{phE}(e)). Considering the short electron
escape depth, the bulk bands observed here are most likely
originated from the FeAs plane below the LaO plane, for which the
calculation indicates its carrier concentration is close to that of
the bulk.\cite{surface} The $k_z$ dispersion suggests that the band
structure quickly becomes bulk-like in the layer just underneath the
surface.

For the bands that do not show much $k_z$ dispersion, they could
either surface or bulk ones in such a quasi-two-dimensional system.
We further explore their reactions to the surface disorder effects
through sodium dosing, as the bulk bands are expected to be more
robust than the surface ones. The sample was exposed to a Na
evaporator at 150 $^{\circ}C$ for 10 seconds, which would put about
1 monolayer of Na atoms on the surface. Our data indicate that the
chemical potential of the system is not changed too much after Na
dosing. Thus, we would expect that additional scattering channels
are introduced on the surface by the Na dosing, which would
dramatically change the surface bands. The data taken at
$\pi$-geometry before and after the dosing were shown in
Fig.~\ref{Na-dose}. Indeed, the bulk $\varpi$ band is barely changed
by Na dosing. Consistently, the surface LaO band ($\lambda_{s1}$ or
$\lambda_{s2}$) disappeared after Na dosing (Fig.~\ref{Na-dose}(a)
and (c)). The $\beta_s$ and $\zetaup_s$ bands disappear as well,
while the $\beta$ band survives the dosing. Since the $\beta$ band
reacts to the Na-dosing so differently from the surface bands, we
attribute the $\beta$ band to the bulk, while the $\beta_s$ and
$\zetaup_s$ bands to the surface FeAs plane. For the $A$ point, only
the $\delta$ band survives the dosing, indicating its bulk character
as shown in Figs.~\ref{Na-dose}(c) and (f).

Due to the surface charge redistribution, calculation shows that the
doping concentrations and band structures of the several layers
close to surface are different from each other and that in bulk. The
surface carrier concentration could be estimated by the Luttinger
volume of the relevant Fermi surfaces. We identify three surface
bands crossing Fermi level around $\Gamma$, which is consistent with
the surface band calculation.\cite{surface} The carrier occupation
of the surface $\beta_s$, $\gamma_s$, $\zetaup_s$ and $\delta_s$
bands is about 4.39~$e^-$ per [Fe$_2$As$_2$] formula unit, smaller
than the calculated 4.89~$e^-$ (based on the calculated Fermi
surface in Ref.\onlinecite{surface}) and expected 6~$e^-$ for the
bulk bands. \cite{DHLuPhC} On the other hand, the counting for the
states of LaO layer gives rise to 0.53 excess electrons per
[La$_2$O$_2$] formula unit, larger than the calculated number of
0.25~$e^-$ (based on the calculated Fermi surface in
Ref.\onlinecite{surface}). The certain difference may be related to
the fact that some bands (thus some Fermi pockets) could be missing
in the experiments.

\begin{figure}[t]
\includegraphics[width=8cm]{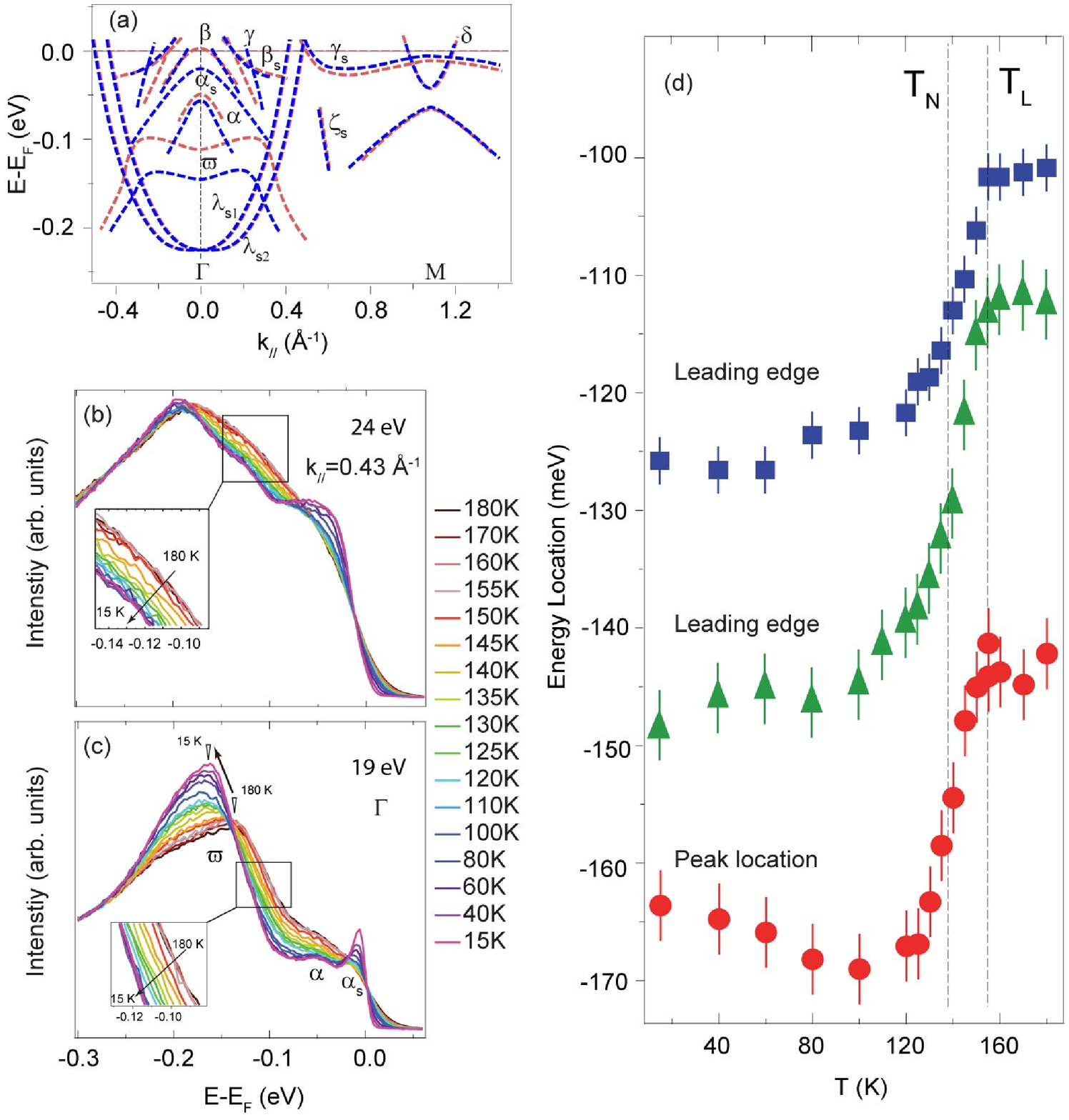}
\caption{(color online) The band shift in LaOFeAs. (a) The
comparison of the band structure at 170~K (red dashed curves) and
10~K (blue dashed curves) reproduced from Fig~\ref{normal} and
Fig~\ref{sdw}. (b)-(c) The temperature dependence of EDC's at (b)
$k_{\varparallel}$=0.43 $\AA^{-1}$ obtained using 24~eV photons and
(c) the $\Gamma$ point obtained using 19~eV photons. The insets of
panel (b) and (c) zooms in the rectangle region to show the leading
edge shift more clearly. (d) The green triangles are the leading
edge locations of the EDC's corresponding to the $\varpiup$ band in
panel (b). The blue squares and the red circles are the leading edge
and peak locations of the EDC's corresponding to the $\varpiup$ band
in panel (c). }\label{shift}
\end{figure}

\begin{figure}[t]
\includegraphics[width=8cm]{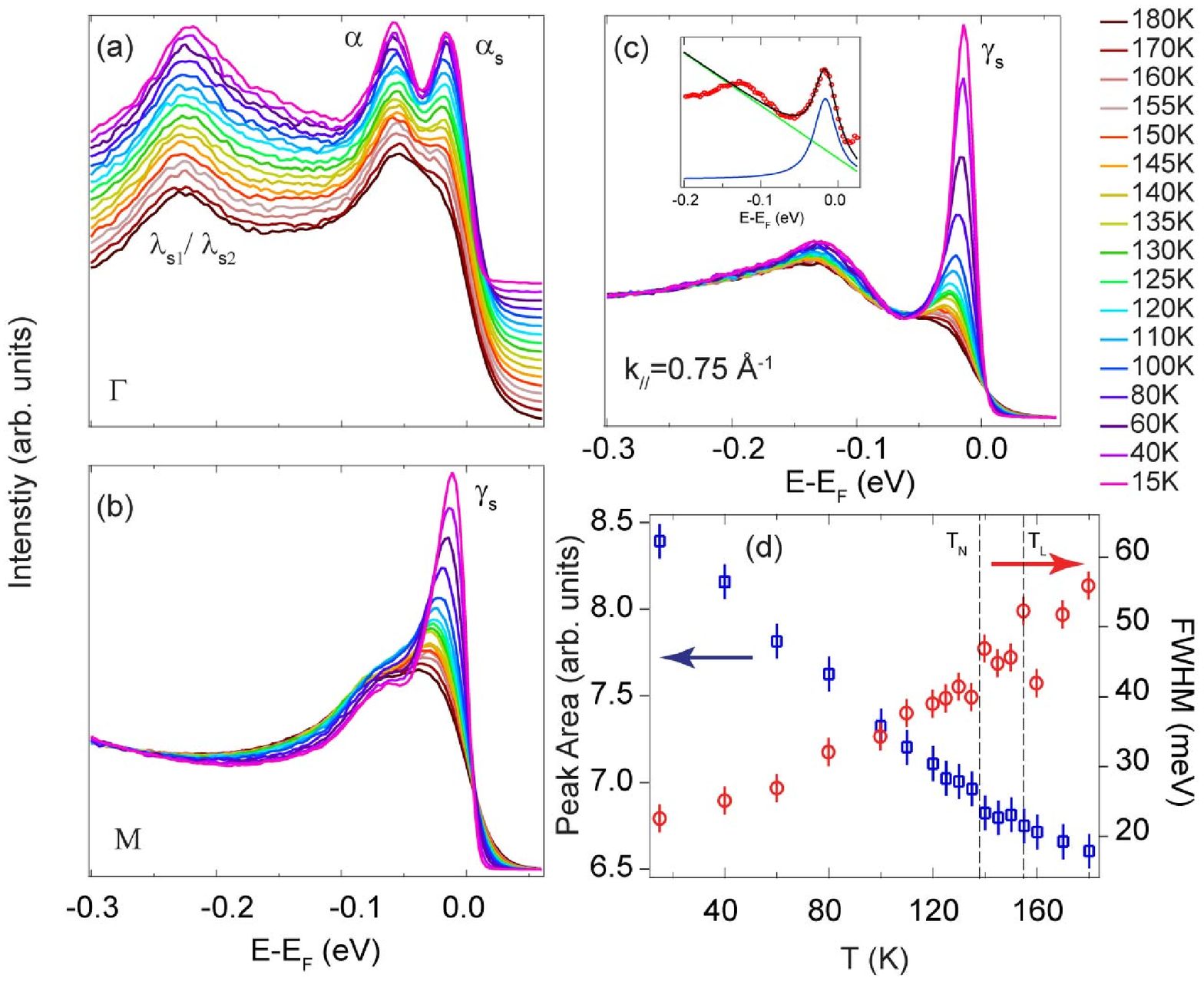}
\caption{(color online) The evolution of the electronic structure in
LaOFeAs. (a)-(c) Temperature dependence of EDC's with momentum
locations at (a) $\Gamma$, (b) $M$ and (c) $k_{\varparallel}$=0.75
$\AA^{-1}$. The inset of panel (c) shows the fitting of the
de-convoluted EDC at 100K. (d) The evolution of the peak area and
full width at half maximum (FWHM) of the EDC's in the panel (c). The
peak area is obtained by integrating the spectral weight of the EDCs
over a window of [-0.065, 0.05]~eV.}\label{peak}
\end{figure}

\subsection{Shifts in bulk band structure correlated with the SDW transition}

The identified bulk bands  allow us to investigate the intrinsic
bulk response of the electronic structure to the phase transitions.
The band structure in the normal state and SDW state are reproduced
and overlaid in Fig.~\ref{shift}(a). The $\varpi$ band clearly
shifts about 30~meV to high energies, which reduces the total energy
of the system effectively and energetically favors the SDW
transition. The temperature dependence of EDC's at
$k_{\varparallel}$=0.43 $\AA^{-1}$ are tracked to examine the
evolution of the $\varpi$ band. As shown in Fig.~\ref{shift}(b),
both the peak position and the leading edge shift about 30 ~meV
towards the high energy from 180~K to 15~K. The shift happens around
structural transition and goes through SDW transition smoothly (the
inset of Fig.~\ref{shift}(b) and the green triangles in
Fig.~\ref{shift}(d)). Since the dispersion of $\varpi$ is concealed
by other bands here, data taken with 19 eV photons are presented in
Fig.~\ref{shift}(c), where the $\varpi$  band is more enhanced due
to certain matrix element effects.  The band shift as much as
25$\pm4$~meV is quantified by both the peak positions (red circles)
and leading edge locations (blue squares) as shown in
Fig.~\ref{shift}(e). Data obtained in different ways all show that
the band shift indeed starts from the structural transition, and
proceeds smoothly across the SDW transition to high energies.

This fact suggests that such band shift is related to the structural
transition. However, the 0.52$\%$ distortion of the lattice could
only account for about 4~meV band shift, which is far too less than
the observation. On the other hand, the band shift could be related
to the magnetic ordering considering its same energy scale as the
magnetic exchange interactions.\cite{LXYang, YZhang, BZhou, CHe}
Therefore, the smoothly evolution of the band shift across the SDW
transition could be interpreted that there is already SDW
fluctuation or short range magnetic order at the structural
transition. Particularly, such a shift of band at the structural
transition temperatures has been observed in NaFeAs before.
\cite{CHe} Furthermore, it has been proved that a short range
magnetic order emerges around the structural transition temperature,
since the band folding due to such an order is observed starting
from the structural transition temperature.\cite{CHe} Therefore, the
band shift in NaFeAs and LaOFeAs can be directly associated with the
SDW fluctuations. Coincidentally, it has been proposed that the
magnetic fluctuations could drive the structural
transition.\cite{JPHu} Our result provides direct support for such a
scenario in the ``1111" series of iron pnictides. Moreover, the
observed shift of bands at high binding energies is consistent with
the reconstruction of the electronic structure in the parent
compounds of ``122" series evolving to the SDW state. \cite{LXYang,
YZhang, MYi, BZhou}. Therefore, our observation of the band shift in
LaOFeAs supports the electronic structure reconstruction as the
mechanism for SDW. It might allude to the existence of an electronic
nematic phase in the ``1111" series between structural and SDW
transition as proposed in Ref.\onlinecite{JPHu}.

\subsection{Drastic temperature dependence of quasiparticles}

To further investigate the influence of the structural and SDW
transitions on electronic structure, the sharp quasiparticle peaks
at low temperatures as shown in Fig~\ref{sdw}(f) are examined as a
function of temperature in Fig.~\ref{peak}. At the $\Gamma$ point
(Fig.~\ref{peak}(a)), a prominent coherent quasiparticle peak
emerges as temperature decreases, suggesting a rapid decrease of
scattering rate. The same behavior could be found at a large portion
of Brillouin zone, such as the $M$ point and the location of
$k_{\varparallel}$=0.75 $\AA^{-1}$ (Figs.~\ref{peak}(b) and
\ref{peak}(c)). The drastic rising of the quasiparticle peaks is
well beyond the thermal sharpening effect at low temperatures. In
order to reveal the relation between this anomalous evolution of the
quasiparticle peaks and the phase transition, we divide the
Fermi-Dirac functions of the temperature dependent EDC's in
Fig.~\ref{peak}(c) and fit the quasiparticle peak with a Lorentz
function and a linear background as shown in the inset of
Fig.~\ref{peak}(c). The quasiparticle peak width which reflects the
scattering rate decreases drastically at low temperatures as
indicated by the red circles in Fig.~\ref{peak}(d). Such decrease of
scattering rate is compatible with that in CeOFeAs, \cite{XJZhouCe}
except that surface and bulk bands were not distinguished there.

The scattering from spin fluctuations could be dramatically
suppressed by the formation of a spin gap below SDW
transition.\cite{SrNeutron, BaNeutron} Consistently, the
quasi-particle weight near $E_F$ increases at low temperature as
shown in Fig.~\ref{peak}(d) (blue squares). Furthermore, the surface
is still a quasi two dimensional system, it could be coupled to the
bulk, for example, through certain scattering processes. Therefore,
the drastic temperature dependence of the surface bands might be
related to the SDW transition in the bulk. However,no noticeable
anomaly for the evolution of the peak width at SDW transitions is
observed. One of the possible relevance is that the magnetic
fluctuations happen even above the structural transition. The
relationship of this peculiar behavior and the phase transitions
needs further studies. Indeed, NMR measurements have verified the
change of the fluctuation around structural transition and its
suppression at low temperatures although there is a drastic
enhancement of the fluctuations around SDW transition. \cite{NaNMR,
LaNMR}

\begin{figure}
\includegraphics[width=8cm]{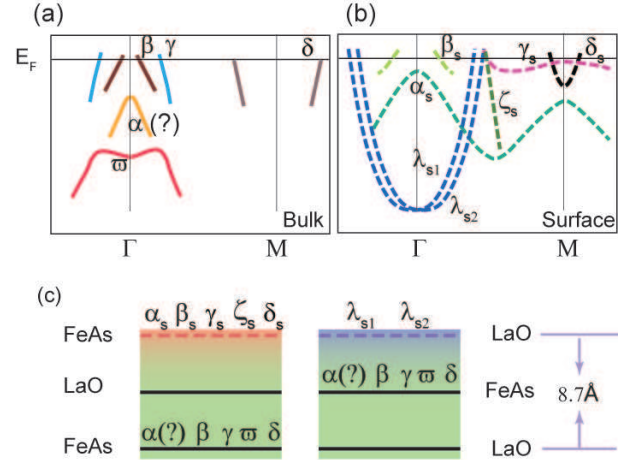}
\caption{(color online) The summary of (a) the surface and (b) bulk
band structures that can be measured. (c) The diagram for the origin
of the measured bands.}\label{summary}
\end{figure}

\section{Summary}

To summarize, we have obtained a comprehensive picture of the
electronic structure in LaOFeAs by distinguishing bulk
(fig.\ref{summary} (a) curves) and surface bands (fig.\ref{summary}
(b)). The origin of the bands is summarized in Fig.\ref{summary} (c)
as well. We note that some certain bands may be missing in the
measurements because of the matrix element effects. Due to the
charge redistribution effect in the ``1111" series, the measured
electronic structure of LaOFeAs is heavily contaminated by the
surface states. Therefore, great caution has to be taken on ARPES
data obtained on this ``1111" series of iron pnictides. For example,
the measured superconducting gap on a surface band might be caused
by the proximity effects, and the strong kink is actually on a
surface band, it does not necessarily suggest strong electron-phonon
interactions in the bulk bands.

Our data show that the large downward shift of the bulk band saves
the total energy of the system and drives the phase transitions. The
onset of this shift at the structural transition temperature further
evidence the existence of a fluctuating magnetic ordering phase that
drives the structural transition. The results in the ``1111" series
are consistent with the observations in the ``122" and ``111" series
of iron pnictides, suggesting the universality of the electronic
response to the structural and magnetic phase transitions. Our
results would help to construct a global picture of the Fe-HTSC
physics.

\textbf{Acknowledgement:} Some preliminary data were taken at beam
line 5-4 of Stanford Synchrotron Radiation Laboratory. This work was
supported by the NSFC, MOST (National Basic Research Program
No.2006CB921300 and 2006CB601002), MOE, and STCSM of China.


\begin{references}

\bibitem{JACS} Y. Kamihara, T. Watanabe, M. Hirano and H. Hosono, J. Am. Chem. Sco. \textbf{130}, 3296 (2008).

\bibitem{XHChenNature} X. H. Chen, T. Wu, G. Wu, R. H. Liu, H. Chen and D. F. Fang, Nature (London) \textbf{453}, 761 (2008).

\bibitem{GFChenCe} G. F. Chen, Z. Li, D. Wu, G. Li, W. Z. Hu, J. Dong, P. Zheng, J. L. Luo and N. L. Wang, Phys. Rev. Lett. \textbf{100}, 247002 (2009).

\bibitem{HosonoNature} Hiroki Takahashi, Kazumi Igawa, Kazunobu Arii, Yoichi Kamihara, Masahiro Hirano, Hideo Hosono, Nature (London) \textbf{453}, 376 (2008).

\bibitem{MXuTh} Min Xu, Fei Chen, Cheng He, Hong-Wei Ou, Jia-Feng Zhao and Dong-Lai Feng, Chem. Mater. \textbf{20}, 7201 (2008).

\bibitem{ChenISO} R. H. Liu, T. Wu, G. Wu, H. Chen, X. F. Wang, Y. L. Xie, J. J. Ying, Y. J. Yan, Q. J. Li, B. C. Shi, W. S. Chu, Z. Y. Wu and X. H. Chen, Nature (London), \textbf{459}, 64 (2008).

\bibitem{RecordTc} Z. A. Ren, W. Lu, J. Yang, W. Yi, X.-L. Shen, Z. Cai, G.-C. Che, X.-L. Dong, L.-L. Sun, F. Zhou and Z.-X. Zhao, Chin. Phys. Lett. \textbf{25}, 2215 (2008).

\bibitem{RecordTc2} C. Wang, L.-J. Li, S. Chi, Z.-W. Zhu, Z. Ren, Y.-K. Li, Y.-T. Wang, X. Lin, Y.-K. Luo, S. Jiang, X.-F. Xu, G.-H. Cao and Z. A. Xu, EPL \textbf{83}, 67006 (2008).

\bibitem{PCDaiNature} Clarina de la Cruz, Q. Huang, J. W. Lynn, Jiying Li, W. Ratcliff II, J. L. Zarestky, H. A. Mook, G. F. Chen, J. L. Luo, N. L. Wang and Pengcheng Dai, Nature, \textbf{453}, 899 (2008).

\bibitem{DHLuNature} D. H. Lu, M. Yi, S.-K. Mo, A. S. Erickson, J. Analytis, J.-H. Chu, D. J. Singh, Z. Hussain, T. H. Geballe, R. R. Fisher and Z.-X. Shen, Nature, 455, 81 (2009).

\bibitem{DHLuPhC} D.H. Lu, M. Yi, S.-K. Mo, J.G. Analytis, J.-H. Chu, A.S. Erickson, D.J. Singh, Z. Hussain, T.H. Geballe, I.R. Fisher and Z.-X. Shen, physica C, \textbf{469}, 452 (2009).

\bibitem{HDing} H. Ding, P. Richard, K. Nakayama, K. Sugawara, T. Arakane, Y. Sekiba, A. Takayama, S. Souma, T. Sato, T. Takahashi, Z. Wang, X. Dai, Z. Fang, G. F. Chen, J. L. Luo and N. L. Wang, Euro. Phys. Lett. \textbf{83}, 47001 (2009).

\bibitem{HDing2} P. Richard, T. Sato, K. Nakayama, S. Souma, T. Takahashi,Y.-M. Xu, G. F. Chen, J. L. Luo, N. L. Wang and H. Ding, phys. Rev. Lett. \textbf{102}, 047003 (2009).

\bibitem{LXYang} L. X. Yang, Y. Zhang, H. W. Ou, J. F. Zhao, D. W. Shen, B. Zhou, J. Wei, F. Chen, M. Xu, C. He, Y. Chen, Z. D. Wang, X. F. Wang, T. Wu, G. Wu, X. H. Chen, M. Arita, K. Shimada, M. Taniguchi, Z. Y. Lu, T. Xiang and D. L. Feng, Phys. Rev. Lett. \textbf{102}, 107002 (2009).

\bibitem{YZhang} Y. Zhang, J. Wei, H. W. Ou, J. F. Zhao, B. Zhou, F. Chen, M. Xu, C. He, G. Wu, H. Chen, M. Arita, K. Shimada, H. Namatame, M. Taniguchi, X. H. Chen, D. L. Feng, Phys. Rev. Lett. \textbf{102}, 127003 (2009).

\bibitem{MYi} M. Yi, D. H. Lu, J. G. Analytis, J.-H. Chu, S.-K. Mo, R.-H. He, M. Hashimoto, R. G. Moore, I. I. Mazin, D. J. Singh,  Z. Hussain, I. R. Fisher and Z.-X. Shen, Phys. Rev. B \textbf{80}, 174510 (2009).

\bibitem{BZhou} Bo Zhou, Yan Zhang, Le-Xian Yang, Min Xu, Cheng He, Fei Chen, Jia-Feng Zhao, Hong-Wei Ou, Jia Wei, Bin-Ping Xie, Tao Wu, Gang Wu, Masashi Arita, Kenya Shimada, Hirofumi Namatame, Masaki Taniguchi, X. H. Chen, and D. L. Feng, Phys. Rev. B \textbf{81}, 155124 (2010).

\bibitem{ZX_PM} M. Yi, D. H. Lu, J. G. Analytis, J.-H. Chu, S.-K. Mo, R.-H. He, R. G. Moore, X. J. Zhou, G. F. Chen, J. L. Luo, N. L. Wang, Z. Hussain, D. J. Singh, I. R. Fisher and Z.-X. Shen, Phys. Rev. B \textbf{80}, 024515 (2009).

\bibitem{Kaminski} Takeshi Kondo, A. F. Santander-Syro, O. Copie, Chang Liu, M. E. Tillman, E. D. Mun, J. Schmalian, S. L. Bud¡¯ko, M. A. Tanatar, P. C. Canfield and A. Kaminski, Phys. Rev. Lett. \textbf{101}, 147003 (2008).

\bibitem{XJZhouCe} Haiyun Liu, G. F. Chen, Wentao Zhang, Lin Zhao, Guodong Liu, T.-L. Xia, Xiaowen Jia, Daixiang Mu, Shanyu Liu, Shaolong He, Yingying Peng, Junfeng He, Zhaoyu Chen, Xiaoli Dong, Jun Zhang, Guiling Wang, Yong Zhu, Zuyan Xu, Chuangtian Chen and X. J. Zhou, arXiv:0912.2838 (unpublished).

\bibitem{surface} H. Eschrig, A. Lankau and K. Koepernik, arXiv:1001.1127 (unpublished).

\bibitem{CHe} C. He, Y. Zhang, B. P. Xie, X. F. Wang, L. X. Yang, B. Zhou, F. Chen, M. Arita, K. Shimada, H. Namatame, M. Taniguchi, X. H. Chen, J. P. Hu, D. L. Feng, arXiv:1001.2981 (unpublished).

\bibitem{Growth} J.-Q. Yan, S. Nandi, J. L. Zarestky, W. Tian, A. Kreyssig, B. Jensen, A. Kracher, K. W. Dennis, R. J. McQueeney, A. I. Goldman, R. W. McCallum, T. A. Lograsso, Appl. Phys. Lett. \textbf{95}, 222504 (2009).

\bibitem{IIMazin} I. I. Mazin, M. D. Johannes, L. Boeri, K. Koepernik and D. J. Singh, Phys. Rev. B \textbf{78}, 085104 (2008).

\bibitem{ZYLu} F. J. Ma and Z.-Y. Lu, Phys. Rev. B \textbf{78}, 033111 (2008).

\bibitem{SinghCalc} D. J. Singh, Phys. Rev. B \textbf{78}, 094511 (2008).

\bibitem{CKim} B. J. Kim, Hosub Jin, S. J. Moon, J.-Y. Kim, B.-G. Park, C. S. Leem, Jaejun Yu, T. W. Noh, C. Kim, S.-J. Oh, J.-H. Park, V. Durairaj, G. Cao, and E. Rotenberg, Phys. Rev. Lett. \textbf{101}, 076402 (2008).

\bibitem{AuSO} H. Cercellier, Y. Fagot-Revurat, B. Kierren, F. Reinert, D. Popovi\ifmmode \acute{c}\else \'{c}\fi{} and D. Malterre, Phys. Rev. B \textbf{70}, 193412 (2004).

\bibitem{HSSO} C. L. Yang, H. T. He, Lu Ding, L. J. Cui, Y. P. Zeng, J. N. Wang and W. K. Ge, Phys. Rev. Lett. \textbf{96}, 186605 (2006).

\bibitem{FChen} Fei Chen, Bo Zhou, Yan Zhang, Jia Wei, Hong-Wei Ou, Jia-Feng Zhao, Cheng He, Qing-Qin Ge, Masashi Arita, Kenya Shimada, Hirofumi Namatame, Masaki Taniguchi, Zhong-Yi Lu, Jiangping Hu, Xiao-Yu Cui and D. L. Feng, Phys. Rev. B \textbf{81}, 014526 (2010).

\bibitem{YZhangOrbital} Y. Zhang, B. Zhou, F. Chen, J. Wei, M. Xu, L. X. Yang, C. Fang, W. F. Tsai, G. H. Cao, Z. A. Xu, M. Arita, H. Hayashi, J. Jiang, H. Iwasawa, C.H. Hong, K. Shimada, H. Namatame, M. Taniguchi, J. P. Hu, D. L. Feng, arXiv:0904.4022 (unpublished).

\bibitem{JPHu} Chen Fang, Hong Yao, Wei-Feng Tsai, JiangPing Hu and Steven A. Kivelson, Phys. Rev. B \textbf{77}, 224509 (2008).

\bibitem{SrNeutron} Jun Zhao, Dao-Xin Yao, Shiliang Li, Tao Hong, Y. Chen, S. Chang, W. Ratcliff II, J.W. Lynn, H. A. Mook, G. F. Chen, J. L. Luo, N. L. Wang, E.W. Carlson, Jiangping Hu and Pengcheng Dai, Phys. Rev. Lett. \textbf{101}, 167203 (2008).

\bibitem{BaNeutron} K. Matan, R. Morinaga, K. Iida and T. J. Sato1, Phys. Rev. B \textbf{79}, 054526 (2009).

\bibitem{NaNMR} Weiqiang Yu, L. Ma, J. Zhang, G. F. Chen, T.-L. Xia, S. Zhang, Y. Hou, arXiv:1004.3581 (unpublished).

\bibitem{LaNMR} Y. Nakai, K. Ishida, Y. Kmihara, M. Hirano, and H. Hosono, J. Phy. Soc. Jpn. 77, 073701 (2008).


\end{references}

\end{document}